\documentclass[letterpaper,english,pra,aps,preprint,nofootinbib,superscriptaddress]{revtex4}%
\usepackage{graphicx}
\usepackage{color}
\usepackage{bm}
\usepackage{amsmath}
\usepackage{amssymb}
\usepackage{amsfonts}

\begin{document}
\title{Reply to Comment on \textquotedblleft Null weak values and the past of a
quantum particle\textquotedblright\ by D.\ Sokolovski}
\author{Q.\ Duprey}
\affiliation{Laboratoire de Physique Th\'{e}orique et Mod\'{e}lisation (CNRS Unit\'{e}
8089), Universit\'{e} de Cergy-Pontoise, 95302 Cergy-Pontoise cedex, France}
\author{A. Matzkin}
\affiliation{Laboratoire de Physique Th\'{e}orique et Mod\'{e}lisation (CNRS Unit\'{e}
8089), Universit\'{e} de Cergy-Pontoise, 95302 Cergy-Pontoise cedex, France}

\begin{abstract}
We discuss the preceding Comment and conclude that the arguments given there
against the relevance of null weak values as representing the absence of a
system property are not compelling. We give an example in which the transition
matrix elements that make the projector weak values vanish are the same ones
that suppress detector clicks in strong measurements. Whether weak values are
taken to account for the past of a quantum system or not depend on general
interpretional commitments of the quantum formalism itself rather than on
peculiarities of the weak measurements framework.

\end{abstract}
\maketitle

\section{\label{intro}Introduction}

The meaning of weak values has been debated since their inception \cite{aav}.
Although the weak measurements framework is entirely derived from standard
quantum mechanical quantities, the axiomatics of quantum theory do not
prescribe any meaning to the weak values. The very basic question is whether
weak values can be taken as a generalized form of eigenvalues, and hence refer
to values taken by properties of the system on par with the values obtained
from projective measurements. A recent focal point of this debate has been the
implications when assessing the past evolution of a quantum particle from weak
values. Vaidman noted \cite{vaidman} that by inserting a small Mach-Zehnder
interferometer (MZI) along the arm of a larger MZI, the particle's presence as
inferred from the weak values was detected inside the nested MZI, but not
along the entrance or exit arms, where the relevant weak values vanished.
Discontinuous trajectories were also noted in a 3 paths interferometer
\cite{A2013JPA}, though seen to depend on the observable chosen to weakly
probe the particle. The paradoxes appearing when employing the "weak trace
criterion" -- Vaidman's suggestion of inferring \cite{vaidman} the presence of
a quantum particle from the traces left on weakly coupled meters -- led
several authors (see Refs [12-23] of \cite{duprey2017} and
\cite{zhou2017,comments,englert} for more recent work) to entirely question
the weak values approach.

Against this backdrop we proposed in our paper \cite{duprey2017}, henceforth
P, an analysis of null weak values. We\ argued that it is consistent to
maintain that vanishing weak values are indicative of the absence of the
corresponding system property in a pre and post-selected system provided (i)
the property is understood as relevant to a transition to the postselected
state and (ii) the quantum system is thought of an extended undulatory entity,
not as a point-like particle. A null weak value of an observable $A$ thus
indicates the local value taken by the property represented by $A$ as the
system evolves and is finally detected in the postselected state.

In the preceding Comment \cite{sokoC}, Sokolovski asserts that even understood
in the liberal sense advocated in P, weak values do not represent physical
properties: they are related to probability amplitudes that are just
computational tools. In the Comment (henceforth C), Sokolovski further
attempts to explain how computational quantities can nevertheless be
determined experimentally from the response of a weak pointer -- this was
indeed one of the main arguments we had given in P in favor of the physical
relevance of vanishing weak values.

In the present Reply, we first examine the novel arguments put forward in the
Comment (Sec.\ II).\ We then give an example comparing vanishing transition
amplitudes as measured in a weak measurement scheme and as inferred from
projective measurements (Sec.\ III). On the basis of this example, we
reexamine in Sec. IV the position outlined in C, and compare it to the weak
trace criterion as well as to our interpretation put forward in P. Our
conclusions given in P remain unchanged.

\section{Weak values and transition amplitudes: physical property or
computational tool?}

In P (Sec. IV.C.2) we had criticized Sokolovski's position \cite{soko}
according to which an experimentally measurable quantity is no more than a
computational tool, on the ground that asserting that an experimentally
measurable quantity was a computational tool was a peculiar position. The main
merit of the Comment, in our view, is to clarify this point.

In C, it is reaffirmed that only strong projective measurements lead to real
paths and that transition amplitudes concern virtual paths and hence as such
they belong to the realm of computational quantities, indicating how certain
terms add up or cancel. Although Sokolovski recognizes that the achievement of
the weak measurement framework is to have discovered a scheme measuring the
response of a weakly perturbed system, weak values are no more, in his view,
than a consequence of perturbation theory. The conclusion is that if no
probabilities are produced, attributing any reality to the transition
amplitudes will lead to unwanted and unecessary paradoxes.

We agree on one point: attributing the property of a weak value to a property
of a localized particle will indeed lead to paradoxes. The particle aspect is
intimately linked to projective measurements that produce probabilities. What
we proposed in P is that the transition amplitudes could be regarded as local
properties of a pre-post selected system understood as a sort of undulatory
entity extended all over space. We did not assert that this was the
\emph{only} consistent approach to account for these experimentally measurable
transition amplitudes, and we had further specified that the choice of the
approach depended on whether one endorsed the assumption linking properties of
a quantum system to projective measurements.

From this point of view, we disagree with the statement made in C assserting
that the approach relying on strong measurements would be the only consistent
way to understand vanishing weak values. We note in particular that Sokolovski
has not shown in C that our proposal was inconsistent. We now turn to a simple
example aimed at emphasizing the link between vanishing transition amplitudes
in strong and weak measurements.

\section{Transition amplitudes with strong measurements: an example}

Let us go back to the 3 path interferometer presented in Fig. 1 of P
(identical to Fig. 1c of C). The weak values along the path are given by
\begin{align}
t  &  =t_{1}\text{ : }\Pi_{E}^{w}=1\qquad\Pi_{F}^{w}=-1\label{ex1}\\
t  &  =t_{2}\text{ : }\Pi_{D}^{w}=1\qquad\Pi_{O}^{w}=0\label{ex2}\\
t  &  =t_{3}\text{ : }\Pi_{E^{\prime}}^{w}=1\qquad\Pi_{F^{\prime}}%
^{w}=-1\label{ex3}\\
t  &  =t_{4}\text{ : }\Pi_{O^{\prime}}^{w}=0, \label{ex4}%
\end{align}
(see Sec. III.C of P, and \cite{A2013JPA} for computational details). The
transition amplitudes at $O\ $and $O^{\prime}$ vanish, $\left\langle \chi
_{f}(t_{2})\right\vert \Pi_{O^{\prime}}\left\vert \psi(t_{2})\right\rangle
=\left\langle \chi_{f}(t_{4})\right\vert \Pi_{O^{\prime}}\left\vert \psi
(t_{4})\right\rangle =0$, explaining why the weakly coupled pointers at $O$
and $O^{\prime}$ do not pick up a shift.

Let us now replace the weakly coupled pointers by pointers having a strong
coupling. The initial state of the pointer located at $X,$ $\left\vert
\varphi_{X}^{{}}(0)\right\rangle $ shifts to $\left\vert \varphi_{X}^{{}%
}(s)\right\rangle $ after the interaction, the strong interaction imposing
$\left\langle \varphi_{X}^{{}}(s)\right\vert \left.  \varphi_{X}^{{}%
}(0)\right\rangle =0$. As in P, $\left\vert \psi\right\rangle $ and
$\left\vert \Psi\right\rangle $ designate the system and total state vector.
Let us start by placing a single strong pointer at any of the positions
$X=E,F,D..$. shown in Fig. 1. The initial state%
\begin{equation}
\left\vert \Psi(t_{i})\right\rangle =\left\vert \psi(t_{i})\right\rangle
\left\vert \varphi_{X}^{{}}(0)\right\rangle
\end{equation}
evolves right after the interaction time $t_{X}$ [corresponding to the
relevant interaction time as per Eqs. (\ref{ex1})-(\ref{ex4})] to
\begin{equation}
\left\vert \Psi(t)\right\rangle =\left\vert \psi_{X}(t)\right\rangle
\left\vert \varphi_{X}^{{}}(s)\right\rangle +\left\vert \psi_{\bar{X}%
}(t)\right\rangle \left\vert \varphi_{X}^{{}}(0)\right\rangle \label{puf}%
\end{equation}
where we use the notation $\left\vert \psi(t)\right\rangle \equiv$ $\sum
_{X}\left\vert \psi_{X}(t)\right\rangle $ as a shorthand equivalent to Eq.
(15) of P (with $\left\vert \psi_{X}(t)\right\rangle =d_{X}(\alpha)\left\vert
m_{\alpha}=X\right\rangle \left\vert \xi_{X}(t)\right\rangle $). $\left\vert
\psi_{\bar{X}}(t)\right\rangle $ is the fraction of the state that evolves
without passing through $X$ and thus does not interact with the pointer. Eq.
(\ref{puf}) is rewritten introducing the projectors $\Pi_{X},$ $\Pi_{\bar{X}}$
as
\begin{equation}
\left\vert \Psi(t)\right\rangle =\Pi_{X}\left\vert \psi(t)\right\rangle
\left\vert \varphi_{X}^{{}}(s)\right\rangle +\Pi_{\bar{X}}\left\vert
\psi(t)\right\rangle \left\vert \varphi_{X}^{{}}(0)\right\rangle
\end{equation}
The system then evolves unitarily ($U$ is the evolution operator) until
$t=t_{f}$ at which point the system is measured and found, say, in state
$\left\vert \chi_{f}\right\rangle .$ With $Q_{X}$ labeling the position of the
pointer, we have at $t=t_{f}$
\begin{equation}
\left\langle Q_{X}\right\vert \left\langle \chi_{f}\right\vert \left.
\Psi(t_{f})\right\rangle =\left\langle \chi_{f}\right\vert U(t_{f},t_{X}%
)\Pi_{X}\left\vert \psi(t_{X})\right\rangle \varphi_{X}^{{}}(Q_{X}%
,s)+\left\langle \chi_{f}\right\vert U(t_{f},t_{X})\Pi_{\bar{X}}\left\vert
\psi(t_{X})\right\rangle \varphi_{X}^{{}}(Q_{X},0). \label{zyx}%
\end{equation}
We see that if a single strongly coupled pointer is inserted at $X,$ it will
undergo a shift and detect the particle with certainty provided the transition
amplitude $\left\langle \chi_{f}\right\vert U(t_{f},t_{X})\Pi_{\bar{X}%
}\left\vert \psi(t_{f})\right\rangle $ vanishes (assuming $\left\langle
\chi_{f}\right\vert U(t_{f},t_{X})\Pi_{X}\left\vert \psi(t_{X})\right\rangle
\neq0$). The transition amplitudes appearing in Eq. (\ref{zyx}) are the same
that appear in the expression of the weak values, Eqs. (\ref{ex1}%
)-(\ref{ex4}).\ Hence a null weak value also implies that a single strongly
coupled pointer will never detect the particle: the property ascription (that
is, giving a value to the observable that is being measured) can be
straightforwardly extended from the strong to the weak coupling case.

\begin{figure}[tb]
	\setlength{\fboxsep}{-45pt}\setlength{\fboxrule}{0pt}\fbox{\includegraphics[scale=0.60]{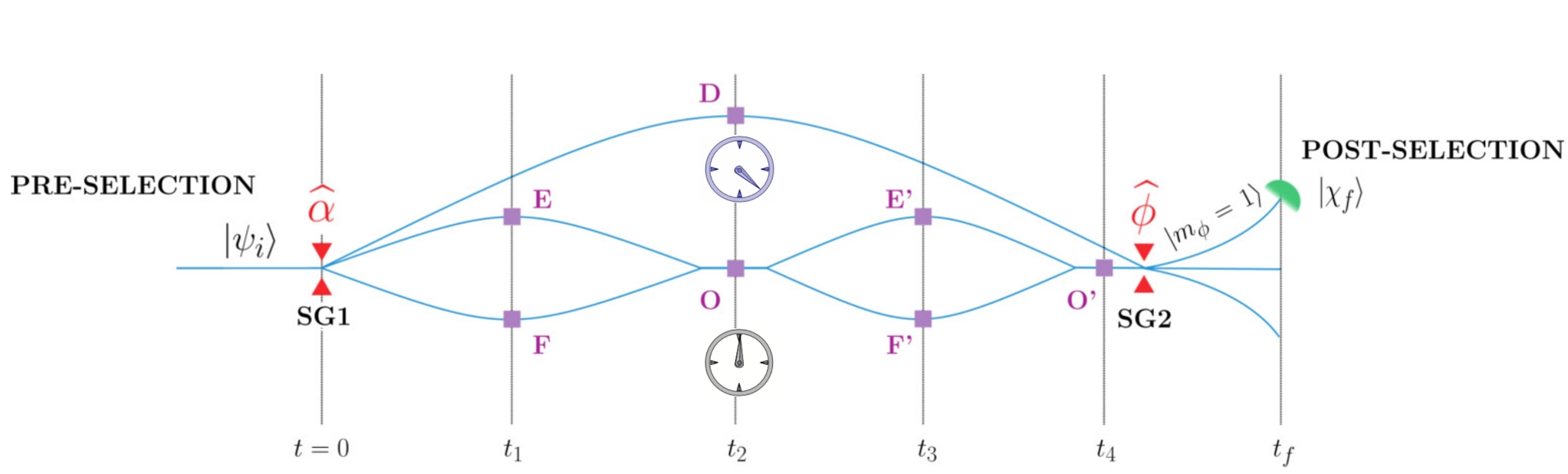}}%
	\caption{The 3 path interferometer for spin-1 particles discussed in P is shown here with strongly coupled pointers placed at $D$ and $O$. The pointer at $D$ always detects the particle, while the pointer at $O$ remains in its ready state. Strong measurements coincide with the weak values (\ref{ex2}) because the transition amplitudes accounting for the strongly coupled pointers are the same ones that enter the definition of the weak values.  }%
	\label{Fig}%
\end{figure}

As an example, assume we place strong pointers at $D$ and $O$ (see Fig. 1). Then the
initial state%
\begin{equation}
\left\vert \Psi(t_{i})\right\rangle =\left\vert \psi(t_{i})\right\rangle
\left\vert \varphi_{D}^{{}}(0)\right\rangle \left\vert \varphi_{O}^{{}%
}(0)\right\rangle
\end{equation}
evolves right after the interaction time $t_{2}$ to
\begin{equation}
\left\vert \Psi(t)\right\rangle =\left\vert \psi_{D}(t)\right\rangle
\left\vert \varphi_{D}^{{}}(s)\right\rangle \left\vert \varphi_{O}^{{}%
}(0)\right\rangle +\left\vert \psi_{O}(t)\right\rangle \left\vert \varphi
_{D}^{{}}(0)\right\rangle \left\vert \varphi_{O}^{{}}(s)\right\rangle
,\label{fonci}%
\end{equation}
that can be rewritten as%
\begin{equation}
\left\vert \Psi(t)\right\rangle =\Pi_{D}\left\vert \psi(t)\right\rangle
\left\vert \varphi_{D}^{{}}(s)\right\rangle \left\vert \varphi_{O}^{{}%
}(0)\right\rangle +\Pi_{O}\left\vert \psi(t)\right\rangle \left\vert
\varphi_{D}^{{}}(0)\right\rangle \left\vert \varphi_{O}^{{}}(s)\right\rangle
.\label{fonci2}%
\end{equation}
The postselected state $\left\vert \chi_{f}\right\rangle $ is chosen so that
$\left\langle \chi_{f}(t_{f})\right\vert U(t_{f},t_{4})\left\vert
\psi_{O^{\prime}}(t_{4})\right\rangle =0$ also implying
\begin{equation}
\left\langle \chi_{f}(t_{f})\right\vert U(t_{f},t_{2})\Pi_{O}\left\vert
\psi(t_{2})\right\rangle =0.\label{gup}%
\end{equation}
When $\left\vert \Psi(t)\right\rangle $ evolves up to $t_{f}$ and the system
is measured and found in $\left\vert \chi_{f}\right\rangle ,$ the final state
becomes%
\begin{align}
\left\langle Q_{D},Q_{O}\right\vert \left\langle \chi_{f}\right\vert \left.
\Psi(t_{f})\right\rangle  & =\left\langle \chi_{f}\right\vert U(t_{f}%
,t_{2})\Pi_{D}\left\vert \psi(t_{2})\right\rangle \varphi_{D}^{{}}%
(Q_{D},s)\varphi_{O}^{{}}(Q_{O},0)\\
& +\label{zrigo}\left\langle \chi_{f}\right\vert U(t_{f},t_{2})\Pi_{O}\left\vert \psi
(t_{2})\right\rangle \varphi_{D}^{{}}(Q_{D},0)\varphi_{O}^{{}}(Q_{O},s).
\end{align}
Since the transition amplitude $\left\langle \chi_{f}\right\vert U(t_{f}%
,t_{2})\Pi_{O}\left\vert \psi(t_{f})\right\rangle $ vanishes, the path
followed by the system is revealed unambiguously by the strongly coupled
pointers. The pointer at $D$ has shifted, while the pointer at $O$ has
remained in the initial state. The system, when found in state $\left\vert
\chi_{f}\right\rangle ,$ has always gone through $D$ and will never be found
at $O$. Note that an additional pointer with a dynamical variable
$Q_{O^{\prime}}$ can be added at $O^{\prime}$ without changing the narrative:
since the transition amplitude $\left\langle \chi_{f}\right\vert U(t_{f}%
,t_{2})\Pi_{O^{\prime}}\left\vert \psi(t_{f})\right\rangle $ vanishes the
extra term appearing in $\left\langle Q_{D},Q_{O},Q_{O^{\prime}}\right\vert
\left\langle \chi_{f}\right\vert \left.  \Psi(t_{f})\right\rangle $ relative
to Eq. (\ref{zrigo}) vanishes and only the single term $\varphi_{D}^{{}}%
(Q_{D},s)\varphi_{O}^{{}}(Q_{O},0)\varphi_{O}^{{}}(Q_{O^{\prime}},0)$ remains.
We conclude that due to the fact that the transition amplitudes are the same
in the strong coupling case and in the weak coupling case examined in P,
although the physics is different, extending the property ascription from the
strong to the weak coupling case is straightforward.

Let us now place strongly coupled pointers at $D,$ $O$, $E^{\prime}$ and
$F^{\prime}$ (see Fig. 2). Shortly after $t=t_{2}$ the quantum state is again given by Eq.
(\ref{fonci}), factored by the ready states $\left\vert \varphi_{E^{\prime}%
}^{{}}(0)\right\rangle \left\vert \varphi_{F^{\prime}}^{{}}(0)\right\rangle .$
However after $t_{3}$ we have%
\begin{align}
\left\vert \Psi(t)\right\rangle  &  =\left\vert \psi_{D}(t)\right\rangle
\left\vert \varphi_{D}^{{}}(s)\right\rangle \left\vert \varphi_{O}^{{}%
}(0)\right\rangle \left\vert \varphi_{E^{\prime}}^{{}}(0)\right\rangle
\left\vert \varphi_{F^{\prime}}^{{}}(0)\right\rangle \nonumber\\
&  +\left(  \left\vert \psi_{E^{\prime}}(t)\right\rangle \left\vert
\varphi_{E^{\prime}}^{{}}(s)\right\rangle \left\vert \varphi_{F^{\prime}}^{{}%
}(0)\right\rangle +\left\vert \psi_{F^{\prime}}(t)\right\rangle \left\vert
\varphi_{E^{\prime}}^{{}}(0)\right\rangle \left\vert \varphi_{F^{\prime}}^{{}%
}(s)\right\rangle \right)  \left\vert \varphi_{D}^{{}}(0)\right\rangle
\left\vert \varphi_{O}^{{}}(s)\right\rangle .
\end{align}
After postselection and measurement of the 4 pointers' positions, we see that
either only $D$ clicks, or alternatively $O$ and $E^{\prime}$, \emph{or }$O$
and $F^{\prime}$ click. The transition amplitude (\ref{gup}) that ensured the
$O$ pointer did not move does not appear anymore. Indeed, the loss of
coherence due to the pointers at $E^{\prime}$ and $F^{\prime}$ being in
different states now correlate a particle detected either at $E^{\prime}$ or
at $F^{\prime}$ with a particle detected at $O.$ We therefore conclude that
that a negative strong measurement coincides with a vanishing weak value
(here, $\Pi_{O}^{w}=0$) if the quantum state is not disturbed by the strong
measurements.\ Otherwise the transition amplitude [here, Eq. (\ref{gup})] that
accounts for vanishing weak values or the lack of clicks in strong
measurements that do not disturb the overall quantum state does not even appear.

\begin{figure}[tb]
	\setlength{\fboxsep}{-45pt}\setlength{\fboxrule}{0pt}\fbox{\includegraphics[scale=0.60]{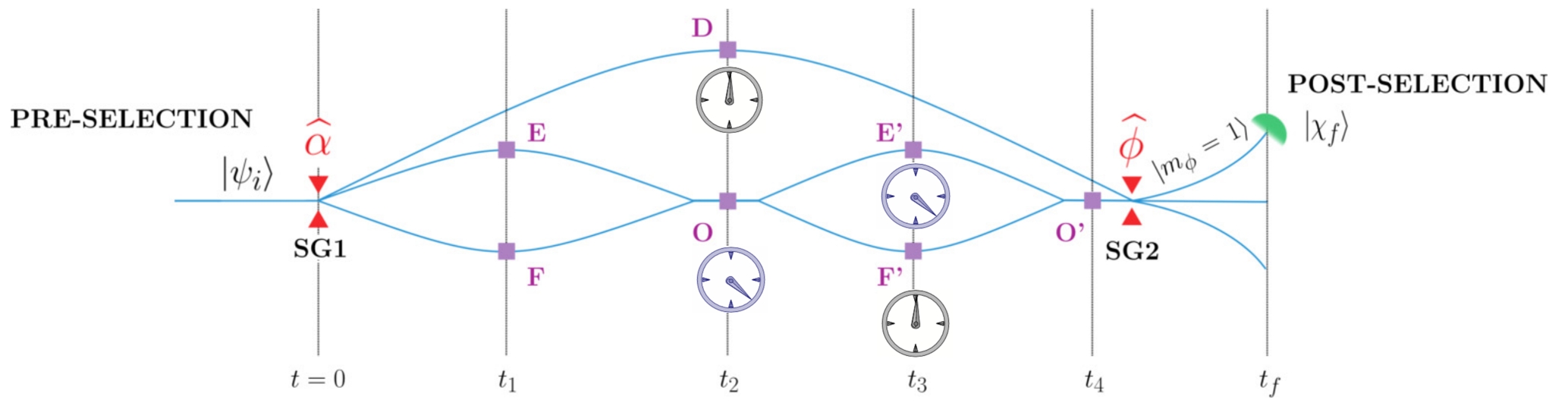}}%
	\caption{Same interferometer displayed in Fig. 1 but shown here with strongly coupled pointers placed at $D$, $O$, $E^{'}$ and $F^{'}$. In the case pictured in the Figure, only the pointers at $O$, and $E^{'}$ click (see the main text for the other possibilities). The strong measurements at $E^{'}$ and $F^{'}$ disturb the quantum state and the transition amplitudes are now different than in the weak case; in particular the vanishing transition amplitude yielding $\Pi_{O}^{w}=0$ does not appear when strong measurements disturb the quantum state.}%
	\label{FigB}%
\end{figure}

\section{Discussion and Conclusion}

The example given above indicates that strong and weak measurements can be set
on the same footing regarding the absence of a property provided the strong
measurements do not disturb the quantum state. The transition amplitudes that
account for the strong pointers motion are the same quantities that enter the
definition of the weak values (\ref{ex1})-(\ref{ex4}). For example in Eq.
(\ref{zrigo}), the $O\ $pointer does not detect the particle because
$\left\langle \chi_{f}(t_{f})\right\vert U(t_{f},t_{2})\Pi_{O}\left\vert
\psi(t_{2})\right\rangle =0$, and from there follows the conclusion that the
particle is only detected along the upper path of Fig.\ 1 (through $D$).\ For
the same reason the weak value $\Pi_{O}^{w}$ vanishes and given the same
configuration (one weakly coupled pointer at $D$, another at $O$), reaching
the same conclusion as in the strong measurement case doesn't seem to be a
problem (although strictly speaking no real pathways are generated).\ Setting
additional weakly coupled pointers along the lower path $E/F\rightarrow
O\rightarrow E^{\prime}/F^{\prime}\rightarrow O^{\prime}$ does not change the
result or the meaning of $\Pi_{O}^{w}$ but the system spatial presence , as
captured by the weakly coupled pointers at $E,F,E^{\prime}F^{\prime}$, appears
as discontinuous. Understood in this manner, extending the concept of property
ascription from the outcomes of strong measurements to weak values is
consistent, provided one keeps in mind that real pathways endowed with a given
probability are not created, and therefore we are not dealing with properties
of a point-like particle. This caveat suffices to avoid dealing with paradoxes.

Extending property ascription based on weak values and transition amplitudes
disagrees with Sokolovski's position interpreting a vanishing transition
amplitude as a simple numerical cancellation without any deeper meaning
(Secs.\ V and VIII of C). Sokolovski further explains that the weakly coupled
pointers motion should be understood as the pointers' reaction to a
perturbation, but this does not entail that the pointer reflects the value of
the property possessed by the system. \ As we have discussed in P, keeping the
standard quantum mechanical assumption requiring strong measurements and real
pathways in order to ascribe properties to a quantum system is consistent. But
then, by this assumption, the possibility to ascribe a value to a system
observable as the system evolves from a preselected to a postselected state is
discarded. We do not see any compelling argument to discard this
possibility.\ Depending on how the quantum formalism is interpreted (in
particular the interplay between the state vector, the measurement problerm
and wave-particle duality) one may feel more comfortable with the standard
approach to quantum properties. On the other hand the fact that a weakly
coupled pointer reacts to the interaction with the system by picking up a
universal quantity related to the transition matrix element to the final
system state, rather than by some arbitrary perturbation depending on the
specifics of the pointer, paves the way for a consistent extension of quantum
properties to pre-post selected systems.

\acknowledgements
Partial support from the Templeton Foundation (Project 57758) is gratefully acknowledged.

\end{document}